# Computation offloading to hardware accelerators in Intel® SGX and Gramine Library OS

Dmitrii Kuvaiskii, Gaurav Kumar, Mona Vij

Intel Labs
2111 NE 25th Ave, Hillsboro, OR 97124
{firstname.lastname}@intel.com

**ABSTRACT**

The Intel® Software Guard Extensions (SGX) technology enables applications to run in an isolated SGX enclave environment, with elevated confidentiality and integrity guarantees. Gramine Library OS facilitates execution of existing unmodified applications in SGX enclaves, requiring only an accompanying manifest file that describes the application's security posture and configuration. However, Intel® SGX is a CPU-only technology, thus Gramine currently supports CPU-only workloads. To enable a broader class of applications that offload computations to hardware accelerators – GPU offload, NIC offload, FPGA offload, TPM communications – Gramine must be augmented with device-backed *mmap* support and generic *ioctl* support. In this paper, we describe the design and implementation of this newly added support, the corresponding changes to the manifest-file syntax and the requisite deep copy algorithm. We evaluate our implementation on Intel® Media SDK workloads and discuss the encountered caveats and limitations. Finally, we outline a use case for the presented *mmap/ioctl* support beyond mere device communication, namely the mechanism to slice the application into the trusted enclave part (where the core application executes) and the untrusted shared-memory part (where insecure shared libraries execute).

## 1. INTRODUCTION

Intel Software Guard Extensions (Intel® SGX) is a hardware technology to create what are known as SGX enclaves – opaque regions of memory isolated from the rest of the environment [1,2]. The code and data executed in SGX enclaves are considered trusted and better protected even against privileged attacks.

To facilitate execution of existing unmodified applications in SGX enclaves, several SGX frameworks were proposed over time [3,4,5]. One of these frameworks is Gramine (formerly "Graphene") – a library OS that allows to "lift-and-shift" arbitrary Linux applications and run them securely in an SGX enclave [6,7]. The application doesn't need to be modified or rebuilt to run under Gramine. To enable the application in Gramine and SGX, it is sufficient to write a so-called *manifest file* that describes the application's configuration, allowed interactions with the untrusted environment, files allowed to be accessed, and so on. Gramine loads the manifest file, validates its authenticity and integrity and uses it to load the application and its dependencies and resolve application requests to the host OS (system calls) at runtime.

However, Intel® SGX is a CPU-only hardware technology. SGX enclaves execute on the CPU, and the SGX-protected code and data reside in CPU caches and (in encrypted form) in RAM. Consequently, Gramine currently implements CPU-only interactions with the untrusted environment. For example, the application running inside a Gramine SGX enclave can perform CPU and memory-intensive tasks, read and write files, send and receive network packets, communicate with other SGX enclaves. But the same application **cannot** offload computations to the GPU (as in the OpenCL/CUDA frameworks), gain access to raw network packets on the NIC (as in the DPDK framework), access TPM for sealing/attestation, or communicate with other devices such as TPUs and FPGAs. To the best of our knowledge, no current SGX framework provides capabilities to access such devices in a generic way. We are aware only of research proof-of-concepts that support accelerator devices in an ad-hoc way [8,9].

This paper proposes a generic way to enable seamless communication with untrusted host devices from within the SGX enclave. To achieve this goal, we note that there are two key interfaces in CPU–device communication: the device-backed *mmap* system call and the *ioctl* system call [10,27]. In a nutshell, device-backed *mmap* system call allows to create untrusted memory regions that are shared between the SGX enclave and an arbitrary host device. The *ioctl* system call provides a generic interface to send an arbitrary request to an arbitrary device, passing arbitrarily complex objects to the device's memory and back to the CPU memory (RAM). Perhaps surprisingly, adding support for these two system calls – together with the classic and already-supported `open`, `read`, `write`, `close` system calls – is sufficient to enable interactions between the trusted SGX enclave and the untrusted host device.

We would like to stress that the **goal** of this work is the *enablement* of communication between SGX enclaves and host devices. *Protecting* this communication from eavesdropping and other attacks is a **non-goal** of this paper. In other words, this paper provides building blocks to create the communication channel; adding integrity checks, encryption, side-channel mitigations, etc. on top of this channel is the responsibility of the application. One example of enhancing the security of this communication channel is the Intel PXP (Protected Xe Path) technology for GPU devices [28].

In the following, we present device-backed *mmap* and generic *ioctl* support in Gramine, as well as the corresponding manifest-file syntax and the *ioctl* deep copy algorithm. This work started as an exploration of GPU support in Gramine SGX enclaves, with the particular focus on Direct Rendering Manager (DRM) *ioctl* standard for GPUs [11] and a set of Intel® Media SDK transcode workloads [12]. However, the ideas and implementations in this paper apply equally well to other kinds of accelerator devices, be it TPUs, NICs, FPGAs, etc. Moreover, we also present a way to slice (partition, or compartmentalize) existing applications into the enclave component and the untrusted components (shared libraries that can be executed in untrusted environment) by re-purposing *ioctl* requests and data structures.

We validated our *mmap* and *ioctl* emulations in Gramine by enabling several Intel® Media SDK workloads. In particular, a sample transcode application could achieve 95% of the original throughput when moved into the SGX enclave using Gramine. During this evaluation, we also identified several caveats and limitations of the device-backed *mmap* and *ioctl* support in SGX enclaves. These caveats are not consequences of our Gramine design or implementation, but rather fundamental mismatches between the traditional CPU-device communication patterns and the two-world Intel® SGX memory model. Fortunately, we were able to find work arounds for all encountered caveats.

## 2. BACKGROUND

This paper describes how to enable communication between Intel® SGX enclaves and host devices using *mmap* and *ioctl* system calls' emulation. Below we give background on the Intel® SGX technology and its interaction with the untrusted environment, as well as background on *mmap* and *ioctl* system calls.

### 2.1 Intel® SGX and ECALL/OCALL interface

Intel® SGX is a hardware technology to create SGX enclaves – opaque regions of memory isolated from the rest of the environment [1,2]. Only the code and data placed in the SGX enclave's memory are considered trusted, whereas the rest of the environment – including privileged software such as the operating system and the hypervisor – is considered untrusted and possibly malicious.

The SGX enclave cannot operate in a completely isolated manner; it must communicate with the untrusted environment to obtain inputs and to pass outputs. To this end, Intel® SGX provides the mechanisms of ECALLs (*enclave calls*) and OCALLs (*outside calls*). In ECALLs, the untrusted runtime places input data in shared untrusted memory and invokes enclave entry, and the enclave picks up this input data. In OCALLs, the enclave copies the output data to untrusted memory and exits, and the untrusted runtime picks up this output data. Note that these mechanisms involve placing data in untrusted memory: the untrusted environment **cannot** access trusted enclave memory, but the enclave **can** access untrusted shared memory [13].

Communication via a predefined set of ECALLs/OCALLs is a classic method of passing data between the SGX enclave and the untrusted environment. Though the enclave may read and write data directly from/into untrusted shared memory, this communication method lacks clear security boundaries and checks, and is thus inadvisable [14].

Every software framework for running SGX enclaves provides an initial set of ECALLs/OCALLs and possibly a means to auto-generate additional, user-specific ECALLs/OCALLs. As one example, Intel® SGX SDK introduces a declarative language called EDL (*Enclave Definition Language*) to describe user-specific ECALLs/OCALLs and a tool called *Edger8r* to produce C glue code from the input EDL file [15]. As another example, Occlum and SCONE have a limited set of predefined ECALLs/OCALLs and **no** means to add user-specific ones [4, 16]. Gramine is similar to Occlum and SCONE in this respect.

The lack of ECALL/OCALL interface extensibility in Gramine is intentional: Gramine runs unmodified applications and emulates the Linux system call layer. Thus, the only supported ECALLs are entry points into the unmodified application: the start-process ECALL and the start-thread ECALL. Similarly, the only supported OCALLs are Linux system calls (more specifically, the minimal required subset of them: `open`, `read`, `write`, `mmap`, `close`, `gettime`, etc.). This restricted set of ECALLs/OCALLs is sufficient for many applications and workloads [17].

## 2.2 *mmap* system call

Applications typically rely on the *mmap* (*memory map*) Linux system call to allocate a memory region that is shared with the hardware accelerator (device). The API of this system call looks like this:

```
void* mmap(void* addr, size_t length,
  int prot, int flags, int fd, off_t offset)
```

The basic function of *mmap* is to allocate a memory region of size `length` in the address space of the Linux process. If `addr` is specified, then the Linux kernel tries to allocate the memory region at this exact address; if `addr` is NULL, then the Linux kernel chooses the starting address as it sees fit. The `prot` argument describes access permissions applied on the memory region: the data in this memory region may be read, written and/or executed.

Unfortunately, the *mmap* system call is overloaded with functionality (tough legacy of early UNIX OSes). In particular, *mmap* is used to allocate memory for three dissimilar cases:

**Anonymous mappings.** The `flags` argument must contain `MAP_ANONYMOUS`; the `fd` and `offset` arguments are ignored. This allocates memory used internally by the process and not reflected in any file or device. E.g., the classic *malloc* function typically uses `mmap(..., MAP_ANONYMOUS, ...)` under the hood. Below is an example of allocating a private read-write memory region of 4KB:

```
void* anon_mapping = mmap(NULL, 4096,
  PROT_READ|PROT_WRITE, MAP_ANONYMOUS, -1, 0)
```

**File-backed mappings.** The `flags` argument must **not** contain `MAP_ANONYMOUS`; the `fd` argument refers to the file descriptor of an opened regular file and the `offset` argument specifies the offset in this opened file. This allocates a memory region that is populated with the contents of the file. By specifying an additional MAP_SHARED flag, *mmap* can be instructed to propagate the updates to this mapping back to the underlying file. File-backed mappings are considered to be a more powerful alternative to classic *read/write* system calls. Below is an example of allocating a read-only not-shared 512B-sized memory region to scan the configuration options of the application:

```
file_fd = open("config.txt", O_RDONLY)
void* file_mapping = mmap(NULL, 512,
  PROT_READ, 0, file_fd, 0)
```

**Device-backed mappings.** The `flags` argument must **not** contain `MAP_ANONYMOUS`; the `fd` argument refers to the file descriptor of an opened device (hardware accelerator); the `offset` argument refers to the device-internal memory address. This allocates a memory region that reflects the internal memory of the device. Typically device-backed mappings specify an additional MAP_SHARED flag, such that all updates by the process are immediately visible to the device and vice versa. Below is an example of allocating a 1MB-sized shared memory region to communicate with the GPU device:

`dev_fd = open("/dev/gpu", O_RDWR)`

`void* device_mapping = mmap(NULL, 1048576, PROT_READ|PROT_WRITE, MAP_SHARED, dev_fd, 0)`

To the best of our knowledge, existing SGX software frameworks support only the first two cases of *mmap*. The third case – which is the focus of this paper – is not supported by any SGX framework (including Gramine).

## 2.3 *ioctl* system call

Some applications rely on the *ioctl* (*input/output control*) Linux system call. This system call is used by the application to communicate with devices such as GPUs, TPUs, FPGAs, hard drives, USB flash drives, etc. The *ioctl* system call is special: in contrast to "normal" system calls like *open* with a fixed and rigid interface, *ioctl* is extensible and is used to transmit arbitrarily complex data structures. Consider the corresponding APIs:

`long open(char* file, int flags, mode_t mode)`

`long ioctl(int fd, int cmd, long arg)`

The "normal" *open* system call has fixed and well-defined arguments: `file` is a string-path, `flags` denote predefined options for opening a file (O_APPEND, O_CREAT, etc.), `mode` specifies read-write-execute permissions of the file. Gramine wraps this system call in a fixed OCALL (*ocall_open*) with the same arguments (`file`, `flags`, `mode`) and copies each of these arguments from enclave memory to untrusted shared memory as part of OCALL execution.

The "special" *ioctl* system call doesn't have a fixed pattern: `cmd` is an arbitrary "magic number" request code (supposedly unique to the communicating device, but there is no standard regulating the request codes), and `arg` is a pointer to an arbitrary data structure. The meaning and the layout of the data structure pointed to by `arg` is dictated by the value in `cmd`. The Linux kernel doesn't pose any restrictions on `arg` or `cmd`. Instead, the Linux kernel forwards cmd and arg to the appropriate device driver, and the device driver parses the data structure pointed to by `arg` depending on the value in `cmd` and its own internal logic.

Given the arbitrarily extensible nature of *ioctl*, it is no surprise that SGX software frameworks support only a limited subset of *ioctl* request codes or do not support *ioctl* at all. For example, Intel® SGX SDK doesn't provide support for *ioctl*; however, users may emulate it by writing their own OCALLs in EDL. As another example, the Occlum framework supports only around ten simple *ioctl* request codes (as of this writing). As a final example, we are not aware of any *ioctl* support in the SCONE framework. To the best of our knowledge, none of the above frameworks allows ioctl-based communication between SGX enclaves and devices.

## 3. *MMAP* SUPPORT IN GRAMINE

As mentioned in Section 2.2, Gramine supports only the first two use cases of the *mmap* system call: anonymous mappings and file-backed mappings. Therefore, in this paper we **extend Gramine with support for device-backed mappings**. This newly added support allows to allocate those memory regions that must be shared with the hardware device in untrusted memory, whereas "normal" anonymous and file-backed memory regions are still allocated in enclave memory (i.e., inside the SGX enclave).

As can be seen in Section 2.2, the only difference between already-supported file-backed mappings and newly supported device-backed mappings is the *origin of file descriptor* (`fd` argument). File-backed mappings use file descriptors of the "file" origin, i.e., they are opened by specifying a path to the regular file ("`config.txt`") in the file system. Device-backed mappings use file descriptors of the "device" origin, i.e., they are opened by specifying a path to the pseudo-file ("`/dev/gpu`") under the special-purpose "`/dev`" directory.

Gramine already supports file descriptors of the "device" origin. In Gramine, all files and devices that will be used by the application must be defined in the so-called Gramine **manifest** file. The manifest file is a simple configuration file written in the standard TOML format [22]. Gramine requires that every application is accompanied by this Gramine-specific manifest file**.**

In particular, to allow communication with a device (e.g., with a GPU) in Gramine, one has to add the following line in the manifest:

```
sgx.allowed_files = "dev:/dev/gpu"
```

Notice the `"dev:"` prefix in the path: this prefix informs Gramine that this is not a regular file but a device. Thus, whenever Gramine intercepts an attempt to open this device from the application, the resulting file descriptor will be marked as "device", and all subsequent system calls that operate on this file descriptor will use the "device" flows.

With the above support for "device" file descriptors in place, implementing device-backed *mmap* is simple: Gramine intercepts `mmap(..., dev_fd, ...)`, recognizes that this *mmap* refers to the device, verifies that `MAP_ANONYMOUS` is unset and that `MAP_SHARED` is set, and allocates a region in untrusted memory via the host-Linux *mmap* invocation.

It is important to note that with the support for device-backed mappings, Gramine effectively splits memory used by the application in two parts: trusted enclave memory and untrusted shared memory. It is crucial for the application to treat shared memory regions as completely untrusted – these memory regions are under the control of a possibly malicious OS kernel. Thus, copying data to untrusted shared memory should be carefully designed to help prevent memory leaks. Similarly, copying data from untrusted shared memory to enclave memory should take into account Time-Of-Check-to-Time-Of-Use (TOCTOU) attacks, rollback attacks, replay attacks, etc. Recall that transparent protection of data in device-backed mappings is a non-goal of this work; the protections must be implemented by the application itself.

## 4. *IOCTL* SUPPORT IN GRAMINE

To date, the Gramine Library OS supports only a limited number of *ioctl* request codes. In this paper, we propose **to extend Gramine with support for arbitrary *ioctl* invocations** with arbitrary request codes and arbitrarily complex data structures. The proposed support has an emphasis on security of *ioctl* system calls and user-friendliness of *ioctl* specification process.

Together with the device-backed *mmap* support, the proposed *ioctl* support enables new classes of applications to run on Gramine and Intel® SGX – AI/ML workloads with GPU/TPU offload, network applications with DPDK processing, edge and embedded applications with FPGA support, etc.

The proposed support also paves the way for practical "slicing" of the original application into the trusted enclave and the untrusted rest of the app. In the "slicing" technique (also called *compartmentalization* or *partitioning*), all communication between the trusted-enclave slice and the untrusted-rest slice is performed via *ioctl* system calls; the required communication patterns are encoded in *ioctl* request codes and data structures.

### 4.1 Requirements for *ioctl* requests

The full spectrum of *ioctl* requests is impossible to know: *ioctl* request codes are not standardized, and software/hardware vendors add new *ioctl* requests with each new version of their device drivers. Conservative estimates name thousands of currently available, unique *ioctl* requests [18].

It would be impractical to hard-code all available *ioctl* request codes and their corresponding data structure layouts into Gramine. First, adding several thousand *ioctl* requests would bloat the Trusted Computing Base (TCB) of Gramine. Second, the hard-coded set of *ioctl* requests would quickly become obsolete, necessitating constant updates of Gramine packages. Third, due to lack of standardization of *ioctl* request codes, some *ioctl* requests will inevitably clash, and Gramine developers will have to make a decision which of the clashing *ioctl* requests to retain and which to remove.

The above considerations lead to the following requirement:

**Requirement 1.** The set of available-to-the-application *ioctl* request codes and corresponding data structures must be flexible and defined on per-application basis.

This requirement implies that Gramine should not hard-code any *ioctl* requests. Instead, each application must define *ioctl* request codes it expects to use, as well as define the corresponding data structures.

Upon application initialization, Gramine should parse the requested-by-the-application *ioctl* request codes and data structures; at runtime, Gramine should only allow those *ioctl* requests that were explicitly defined by the application. In other words, Gramine should not hard-code any *ioctl* requests but only provide the means for the application to define *ioctl* requests and be able to parse and execute them.

Fortunately, Gramine already requires every application to be accompanied by a manifest file. It is only natural to extend the manifest with additional configuration parameters for *ioctl* request codes and *ioctl* data structures.

There is another issue that needs solving. Any *ioctl* system call must be emulated via the OCALL mechanism (recall that there is no other more secure way for the SGX enclave to communicate with the untrusted environment, and *ioctl* requests are no exception). Since the application invokes *ioctl* from within the enclave, the data structure object passed to *ioctl* through the `arg` argument also resides in the enclave. Hence, this data structure object must be copied to the untrusted shared memory during *ioctl* handling. Moreover, many *ioctl* requests update the provided data structure in-place before passing control back to the application; hence, the data structure object must be also copied from untrusted shared memory back to the enclave memory.

But *ioctl* data structures may be arbitrarily complex: they can have nested objects (i.e., pointers to other data structures), and the level of nesting can be arbitrary. They can also represent arrays, linked lists, binary trees, and other complex structures. In other words, *ioctl* data structures are not necessarily flat contiguous memory regions, but typically a set of non-contiguous memory regions chained together via pointers and offsets.

**Requirement 2.** Definitions of *ioctl* data structures in the manifest file must be flexible: they must embrace C-style arrays and pointers, allowing for complex nested data structures such as linked lists.

As mentioned above, *ioctls* are emulated via OCALLs, which means that *ioctl* data structures are copied out of the enclave to untrusted shared memory and then back (if required). However, since ioctl data structures can span non-contiguous memory regions, it is not sufficient to perform a **shallow copy** – this would only copy the first of the memory regions, without following pointers to other memory regions [19,20]. For example, a shallow copy of a linked list would only copy the head of the list (with a meaningless pointer to the next list item).

Instead, *ioctl* data structures must be copied via **deep copy**. In this mechanism, all encountered pointers are resolved by (1) copying the pointed-to memory region and (2) rewiring the pointer to the newly copied memory region [19,20]. This pointer-resolving process must happen recursively until all regions of the data structure are copied. (In this work, we assume that there are no cyclic dependencies in the data structure object, i.e., all *ioctl* data structures are representable as trees of memory regions).

Moreover, some parts of *ioctl* data structures must be copied only in one direction: out of the enclave to untrusted memory (OCALL outputs) or from untrusted memory inside the enclave (OCALL inputs). For example, an *ioctl* that queries a value of some GPU parameter should have the data structure's field `param_name` marked as out-of-enclave and the field `param_value` marked as into-enclave. Also, some parts of *ioctl* data structures must not be copied at all – for example, many structures have padding bytes. These padding bytes should not be copied outside the enclave nor inside of it, according to the SGX threat model and the principle of least privilege (otherwise it could introduce enclave memory leaks or unexpected malicious data).

Finally, sometimes ioctl data structure's pointer fields already point to untrusted memory (for example, to the device-backed mapping described in Section 3). In this case, there is no sense in following this pointer and copying its memory region to untrusted memory again. Instead, the pointer must not be rewired but simply kept as-is.

To summarize, *ioctl* data structures must be **selectively deep-copied**:

**Requirement 3.** Definitions of *ioctl* data structures must provide a hint whether parts of the data structure should be copied outside the enclave, inside of it, in both directions, or not copied at all. At runtime,

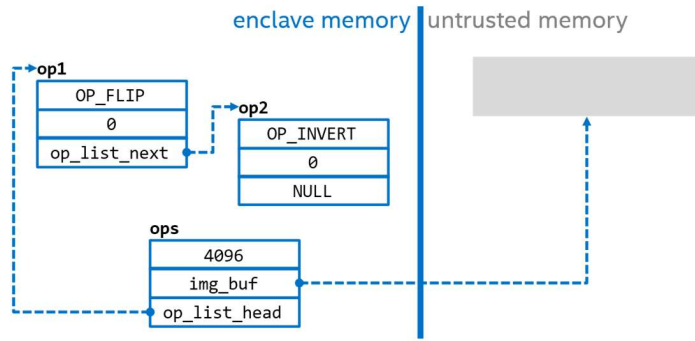

```
struct op {
  int      operation;
  int64_t  seconds_spent;
  struct op* op_list_next;
}
struct ops_for_gpu {
  size_t   img_buf_size;
  void*    img_buf;
  struct op* op_list_head;
}

struct op2 = {OP_INVERT, 0, NULL};
struct op1 = {OP_FLIP, 0, &op2};
struct ops_for_gpu ops = {.., &op1};
ioctl(gpu_fd, 0xc0007102, &ops);
```

**Figure 1.** Example of *ioctl* that instructs the GPU device to perform a series of operations on the image in the untrusted-memory buffer.

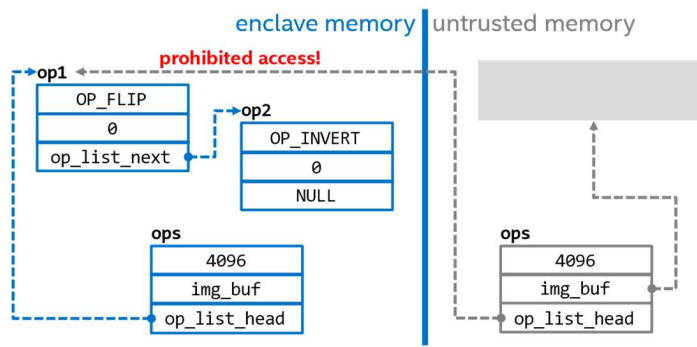

**Figure 2.** Performing a shallow copy leaves a wrong pointer.

*ioctl* emulation must correctly perform selective deep copy of the data structure according to this definition.

The below (contrived) example details the problem of shallow copy during *ioctl* emulation and the expected semantics of the selective deep copy. This hypothetical *ioctl* instructs the GPU device to perform a series of operations on the image in the untrusted-memory buffer and report the time spent on each operation after the GPU is done.

Figure 1 shows the example of this hypothetical *ioctl* in the application code and the corresponding layout of the *ioctl* data structure objects in enclave memory. Notice how `img_buf` pointer field points to the buffer in untrusted memory. Also notice how two operations (to flip the image and then to invert it) form a simple linked list. Finally, pay attention to `seconds_spent` field – it should be updated when the *ioctl* is finished; consequently, there is no sense in copying its value out of the enclave during *ioctl* preparation.

Figure 2 highlights the problem of performing a shallow copy on the *ioctl* data structure. Since shallow copy doesn't follow pointers, the `op_list_head` pointer value is copied as-is and continues pointing to the in-enclave `op1` object. Accessing enclave memory is prohibited when executing the host *ioctl* system call outside of the enclave, thus such an *ioctl* invocation fails.

Figure 3 shows the correct deep copy of this complex *ioctl* data structure. This implementation follows all non-NULL pointers and iteratively copies each new object to untrusted memory. Note the selective nature of this copying: the `seconds_spent` fields are not copied out of the enclave but instead replaced with dummy "0" values in the copies. Also note that pointers in the copied objects are rewired to point to untrusted-memory copies of objects. Finally, a subtle detail is that `img_buf` pointer is not rewired but rather copied as-is (by value).

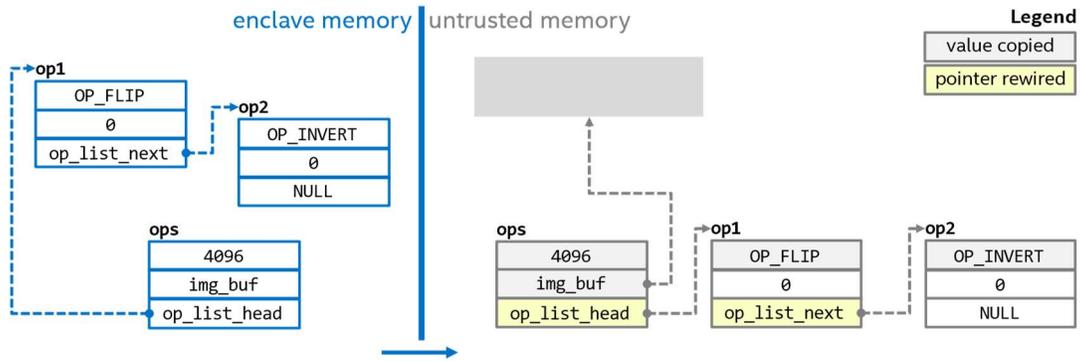

**Figure 3.** Selective deep copy of *ioctl* data structure (from enclave to untrusted memory).

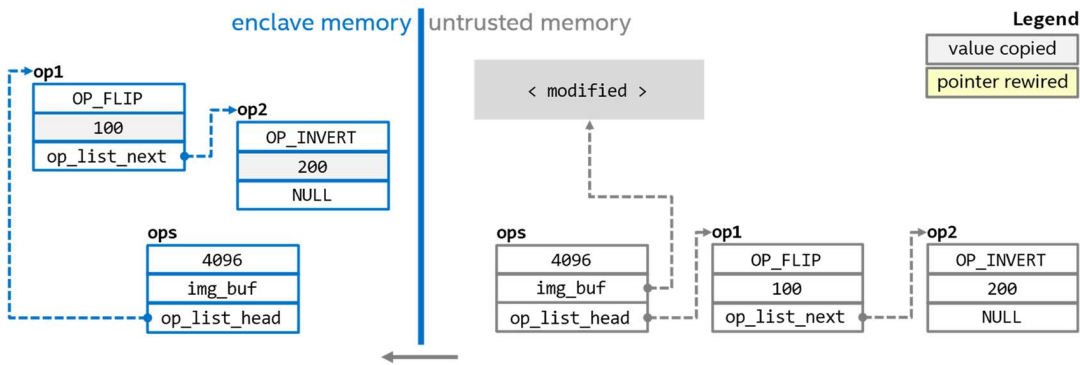

**Figure 4.** Selective deep copy of *ioctl* data structure (from untrusted memory back to enclave).

After this preparation step of copying the objects out of the enclave to untrusted shared memory, the untrusted runtime can execute the actual host *ioctl* system call, and the GPU will start processing the image. When the GPU is done with processing, the host *ioctl* system call returns, and the untrusted runtime must perform the final step: copying the updated fields of the objects back inside the enclave.

Figure 4 shows the reverse deep copy for our example. The actual *ioctl* execution updates the `seconds_spent` fields of `op1` and `op2` objects. These updated fields must be propagated back to the in-enclave objects. To this end, the reverse deep copy selects only these two fields for back-copy. Also notice how the in-enclave `ops` object points to the same image buffer, but now this image buffer is updated with *ioctl*-requested transformations. In the end, the *ioctl* emulation results in the exact same observable behavior to the application as the original non-SGX execution.

There is one more unique aspect about *ioctl* data structures. In contrast to classic ECALL/OCALL data structures, *ioctl* data structures cannot be amended for our SGX needs. In other words, the layouts of *ioctl* data structures passed to/from the SGX enclave are set in stone. As one example, some Direct Rendering Manager (DRM) *ioctl* data structures – used in CPU-GPU communication – specify array sizes not in bytes but in other units of measurement (typically in 32-bit ints). As another example, other DRM *ioctl* data structures specify array sizes adjusted by a certain value (typically to offset the "header" part of these arrays). Also, many *ioctl* data structures must be placed at specific alignments (typically at memory page granularity). Finally, some complex *ioctl* data structures may have different layouts depending on one of their fields (for example, the same DRM *ioctl* data structure may represent a GPU engine-class extension or a GPU engine-load-balancer extension or some other user-defined extension, all based on the value of the `param` field).

These "peculiarities" of existing *ioctl* data structures must be taken into account, since there is no way to fix them for our deep-copy purposes. Instead, our

```
struct op {
  int       operation;
  int64_t   seconds_spent;
  struct op* op_list_next;
}
struct ops_for_gpu {
  size_t    img_buf_size;
  void*     img_buf;
  struct op* op_list_head;
}

struct op2 = {OP_INVERT, 0, NULL};
struct op1 = {OP_FLIP, 0, &op2};
struct ops_for_gpu ops = {.., &op1};
ioctl(gpu_fd, 0xc0007102, &ops);
```

**Manifest file (in TOML syntax)**

```
allowed_ioctls.io1.request = 0xc0007102
allowed_ioctls.io1.struct  = "ops_for_gpu"

ioctl_structs.ops_for_gpu = [
  {name="img_buf_size", size=8, type="out"},
  {name="img_buf", size=8, type="out"},
  {name="op", ptr=[
     {name="operation", size=4, type="out"},
     {name="seconds_spent", size=8, type="in"},
     {name="op_list_next", ptr="op"}
  ]}
]
```

**Figure 5.** Definition of an example *ioctl* request with the corresponding data structure in TOML syntax.

proposed deep copy implementation must account for these peculiarities and provide additional syntax to work around them. Allowing for these peculiarities is **not** implemented in any of the existing deep-copy solutions we are aware of.

**Requirement 4.** Existing *ioctl* data structures have peculiar memory layout patterns that must be taken into account by the *ioctl* definition syntax and by the deep-copy implementation.

In particular, the syntax and the implementation must allow for arbitrary units of measurement, adjustments to memory region sizes, alignments of copied memory regions, and selective copying of memory regions based on values in ioctl data structure's fields.

## 4.2 Considered (and rejected) solutions

The main problem that we solve is marshalling (serializing) complex nested data structures during copy in and out of the enclave. There are multiple solutions for data serialization, including those solutions that support deep copying.

The first solution we considered is the Enclave Definition Language (EDL) which is used by Intel® SGX SDK [15]. EDL is written in a C-like syntax and is used to define the ECALL/OCALL interfaces and data structures passed in/out of the enclave. However, EDL has several disadvantages that make it impractical for our *ioctl* emulation: (1) EDL syntax does not support deep copying, (2) EDL syntax is too verbose and hard to parse, (3) EDL syntax cannot be easily mapped to the TOML syntax in which the rest of Gramine manifest is written. For these reasons, we didn't investigate using EDL for descriptions of *ioctl* data structures. However, we should note that EDL design and choice of keywords affected the implementation of our *ioctl* syntax.

The second solution is to use some standard Interface Definition Language (IDL) like Google Protocol Buffers IDL [21]. Such IDLs typically support deep copying, however, they have the following disadvantages: (1) their syntax is not compatible with the TOML syntax, (2) they are tailored for passing data over the network, (3) they do not have the syntax to account of peculiarities in existing *ioctl* data structures. The second drawback leads to unnecessary complications in the resulting IDL parser implementation (need to convert between network formats, feature overload). The third drawback precludes using IDL languages for our specific *ioctl* use case.

Finally, we are not aware of any TOML-based deep-copy IDL.

Given that there is no existing solution that would satisfy our requirements, we decided to implement our own minimal IDL that borrows many ideas from Intel® SGX SDK's EDL but differs from it in three main aspects: (1) our IDL is written using TOML syntax, (2) our IDL supports deep copy, (3) our IDL natively supports all the peculiarities of existing *ioctl* data structures.

## 4.3 Interface Definition Language for *ioctl*s

Our minimal IDL uses the TOML syntax [22]. TOML is a plaintext file format for configuration files (similar to .INI file format). However, TOML supports two powerful data types – array and table – that allow to define complex structures. Thus, TOML is a good fit for definitions of arbitrarily complex *ioctl* data structures.

We designed our IDL based on the three requirements outlined above. Figure 5 contains an example that explains the main aspects of our TOML-based *ioctl* definitions. In this example, the *ioctl* request itself is defined via the `allowed_ioctls.io1` TOML table. The `io1` key is just a unique name for this particular *ioctl* request. The application and the untrusted host actually identify this ioctl request by the magic number in the `io1.request` field.

The `io1` request uses the `ops_for_gpu` data structure (using a reference by name to the corresponding item in the `ioctl_structs` TOML list). This structure is defined as a combination of TOML inline tables (with the syntax `{key1=value1, key2=value2, ...}`) and TOML inline arrays (with the syntax `[item1, item2, ...]`).

The easiest way to think about this data structure representation is in terms of *memory regions* and *memory sub-regions*. Memory regions are non-contiguous (they represent separate objects in memory). Each memory region contains several contiguous sub-regions (they represent fields of one object). Memory regions are enclosed by TOML's `[]` arrays whereas sub-regions are enclosed by TOML's `{}` tables.

Going through the example in Figure 5, we can read the definition of `ops_for_gpu` as follows:

1. Start with the first memory region. This memory region starts at the address in enclave memory pointed to by ioctl's `arg` argument. It contains three sub-regions.
2. Examine the first sub-region (named `img_buf_size`). Copy 8 bytes out of the enclave to untrusted memory.
3. Examine the second sub-region (at offset 8 bytes due to the first sub-region's size; named `img_buf`). Copy 8 bytes out of the enclave to untrusted memory. Note that this is a pointer to the image buffer in untrusted memory – we must copy this pointer as-is (without rewiring) instead of performing a deep copy of the pointed-to buffer.
4. Examine the third sub-region (at offset 16 bytes due to the previous two sub-regions' sizes; named op). Notice that the name of this sub-region is different from the C struct's field name – the names do not need to match, and the name in the *ioctl* definition is local to this *ioctl* definition only. This third sub-region is a pointer to another memory region. Thus, the deep-copy algorithm must follow this pointer and start examining and copying this pointed-to memory region. Also, this pointer itself must be rewired to point to the newly copied memory region.
5. Jump to the next memory region. This memory region starts at the address in enclave memory pointed to by the op pointer from the previous step. It contains three sub-regions.
6. Examine the three sub-regions. Copy 4 bytes of the first sub-region out of the enclave. **Skip** copying the second sub-region because of its type (`"in"`). The third sub-region is again a pointer, so perform a deep copy of the pointed-to memory region (which is defined by the name op) and rewire the pointer itself (unless the pointer is NULL).
7. Repeat steps 5-6 until there are no new memory regions to copy.
8. When the deep copy is complete, the OCALL is performed. The untrusted runtime uses the deep-copied objects in untrusted memory to execute the host *ioctl*. The deep-copied objects in untrusted memory are updated after the *ioctl* is finished.
9. Examine the whole `ops_for_gpu` definition again, in the same manner as in steps 1-7. But now search for sub-regions with the type `"in"` and ignore all others. When such sub-regions are found, copy the corresponding contents of the (updated) objects in untrusted memory back inside the enclave. In the example, the sub-regions to copy back inside the enclave are named `seconds_spent`.

To satisfy **Requirement 4**, the proposed TOML syntax contains several more keywords to work around the peculiarities of some existing *ioctl* data structures (mostly encountered in the DRM *ioctl* standard). Below is the complete definition of our TOML-based IDL:

- `name` is an optional name for the sub-region; mainly used to find length-specifying fields.
- `align` is an optional alignment of the memory region; may be specified only in the first sub-region of a memory region (all other sub-regions are contiguous with the first sub-region, so specifying their alignment doesn't make sense).
- `size` is a mandatory size of this sub-region. The `size` field may be a string with the name of another field that contains the size value or an integer with the constant size measured in `units` (default unit is 1 byte; also see below). For example, `size="strlen"` denotes a size field that will be calculated dynamically during *ioctl* deep copy based on the sub-region named `strlen`, whereas `size=16` denotes a sub-region of size 16 bytes. Note that for `ptr` sub-regions, the `size` field has a different meaning: it denotes the number of adjacent memory regions (in other words, it denotes the number of items in the `ptr` array).
- `unit` is an optional unit of measurement for `size`. It is 1 byte by default. Unit of measurement must be a constant integer. For example, `size="strlen"` and `unit=2` denote a wide-char string (where each character is 2 bytes long) of a dynamically calculated length.
- `adjust` is an optional integer adjustment for `size`. It is 0 bytes by default. This field must be a constant (possibly negative) integer. For example, `adjust=-8` and `size=12` results in a total size of 4 bytes.
- `type=["none" | "out" | "in" | "inout"]` is an optional direction of copy for this sub-region. For example, `type="out"` denotes a sub-region to be copied out of the enclave to untrusted memory. The default value is none which is useful for e.g. padding of structs. This field may be omitted if the `ptr` field is specified for this sub-region (pointer sub-regions contain the pointer value which will be rewired to point to untrusted memory anyway).
- `ptr=[another memory region]` or `ptr="another-memory-region"` specifies a pointer to another, nested memory region. This field is required when describing complex *ioctl* data structures. Such pointer memory region always has the implicit size of 8 bytes, and the pointer value is always rewired to the memory region in untrusted memory (containing a copied-out nested memory region). If `ptr` is specified together with `size`, it describes not just a pointer but an array of these memory regions.
- `onlyif="simple boolean expression"` allows to condition the sub-region based on a boolean expression. The only currently supported formats of expressions are `token1==token2`, `token1!=token2`, `token1&=token2` (all bits of `token2` are set in `token1`) and `token1|=token2` (at least one bit of `token2` is set in `token1`), where `token1` and `token2` may be constant integers or sub-region names.

### 4.4 Deep copy algorithm

The deep copy algorithm for ioctl request emulation works in four steps: (1) iterate through all memory regions and collect all the corresponding memory sub-regions in one list, (2) selectively copy collected sub-regions from enclave memory to untrusted shared memory and rewire all pointers, (3) supply untrusted-memory copies to the actual host-level *ioctl* execution, (4) selectively copy sub-regions with *ioctl* results from untrusted memory back into enclave memory.

Step 3 is straight-forward, whereas steps 1, 2 and 4 actually implement our deep copy algorithm. These steps (phases) are detailed in pseudo-code in Listings 1, 2, 3 correspondingly.

The first phase is described in Listing 1. Let's describe its more subtle details.

First, recall that the *ioctl* data structure to copy is arbitrarily nested, i.e., the data structure consists of several memory regions. Each memory region consists of several contiguous sub-regions. Thus, the deep copy algorithm must iterate through all memory regions, starting from the root memory region

```
1   root_mem_region = {desc = root_toml_array, addr = ioctl arg}
2   mem_regions = queue(root_mem_region)
3   sub_regions = queue(<empty>)

4   while mem_regions is not empty:
5       current_mem_region = mem_regions.pop()
6       offset_in_mem_region = 0

7       for each sub_region in current_mem_region:
8           if exists sub_region["onlyif"]:
9               skip = evaluate_condition(sub_region["onlyif"])
10              if skip == true:
11                  continue

12          if sub_region["size"].is_string:
13              size_sub_region = sub_regions.find_by_name(sub_region["size"])
14              sub_region.size = read_sub_region(size_sub_region["size"])
15          else:
16              sub_region.size = sub_region["size"].as_integer

17          sub_region.size = sub_region.size * sub_region["unit"]
18          sub_region.size = sub_region.size + sub_region["adjust"]
19          sub_region.addr = current_mem_region.addr + offset_in_mem_region
20          sub_region.align = sub_region["align"]
21          sub_region.type  = sub_region["type"]

22          if exists sub_region["ptr"]:
23              if sub_region["ptr"].is_string:
24                  next_mr_desc = mem_regions.find_by_name(sub_region["ptr"])
25              else:
26                  next_mr_desc = parse_toml_array(sub_region["ptr"])

27              for i = 1..sub_region.size:
28                  next_mr_addr = follow_pointer(sub_region.addr, i)
29                  next_mem_region = {desc = next_mr_desc, addr = next_mr_addr}
30                  mem_regions.push(next_mem_region)

31              sub_region.rewire_pointer = true
32              sub_region.size = 8

33          sub_regions.push(sub_region)
34          offset_in_mem_region = offset_in_mem_region + sub_region.size
```

**Listing 1.** Phase 1 of the deep copy algorithm for *ioctl* request emulation: collecting memory sub-regions. Note that "mr" stands for "memory region".

(defined at build time by the root TOML array corresponding to this *ioctl* request and at run time – by the *ioctl*'s arg argument). Since we need a *selective* deep copy, we shouldn't collect-and-copy memory regions per se, but rather their corresponding sub-regions.

Recall that *ioctl* data structures must be acyclic (i.e., should be representable as trees). The algorithm in Listing 1 assumes this property, otherwise it would go into an infinite loop. The actual implementation allows deep copying only to a predefined level of nesting and errors out if this level is exceeded.

```
1  for each sub_region in sub_regions:
2      align_sub_region_in_untrusted_mem(sub_region.align)

3      if sub_region.rewire_pointer == true:
4          rewire_sub_region_to_untrusted_mem(sub_region)
5      else if sub_region.type == "out" or sub_region.type == "inout":
6          copy_sub_region_to_untrusted_mem(sub_region.addr, sub_region.size)
7      else:
8          zero_sub_region_in_untrusted_mem(sub_region.size)
```

**Listing 2.** Phase 2 of the deep copy algorithm for *ioctl* request emulation: selective copy of collected memory sub-regions from enclave memory to untrusted shared memory.

```
1  for each sub_region in sub_regions:
2      if sub_region.type == "in" or sub_region.type == "inout":
3          copy_sub_region_to_enclave_mem(sub_region.addr, sub_region.size)
4      else:
5          jump_over_sub_region_in_enclave_mem(sub_region.size)
```

**Listing 3.** Phase 3 of the deep copy algorithm for *ioctl* request emulation: selective copy of collected memory sub-regions from untrusted shared memory back to enclave memory.

The algorithm must use Breadth-First Search (BFS) to correctly collect memory regions and sub-regions. This is to allow dynamic calculation of sizes and pointed-to regions for "size" and "ptr" string keywords: typically, the size of the next memory region is defined by one of the sub-regions in the previous memory region. Doing a Deep-First Search (DFS) could lead to situations where the size of the next memory region cannot be identified because the corresponding "size" sub-region hasn't been yet encountered. Careful reader can notice the use of FIFO queues and typical BFS statements on Lines 2-5 and 7, 30, 33.

Some sub-regions may be conditioned by "onlyif" clauses. When detecting such sub-regions, the algorithm evaluates the condition at run time, and if the condition is not satisfied, it skips this particular sub-region (Lines 8-11). This allows for different runt-time representations of the same *ioctl* data structure.

For each sub-region, the algorithm must parse the TOML definition of the sub-region, calculate the run-time parameters of the sub-region, and append it to the queue of sub-regions-to-copy (Lines 12-22 and Line 33). The "size" parameter may be a hard-coded integer, but may be a reference to another sub-region that contains the size value – in this case the value must be read at run-time from this referenced sub-region (Lines 12-14). The size of the sub-region may be optionally adjusted by "unit" and "adjust" parameters, that can only be read as hard-coded integers in the TOML definition (Lines 17-18).

Sub-regions with the "ptr" parameter are special: these are pointers to nested memory regions in this *ioctl* data structure. The pointed-to memory regions are either defined by an already-encountered pointer sub-region (Lines 23-24) or by a new TOML inline array (Lines 25-26). In either case, the new memory region is added to the queue of memory regions, with the corresponding enclave address of this new memory region identified via following the pointer sub-region (Lines 28-30). Also note that for "ptr" sub-regions, the size field has a different meaning: it is not the size of the sub-region in bytes but rather the number of memory regions to copy (Line 27). This syntax allows to define complex-struct arrays. Finally note that the "ptr" sub-region itself is updated to reflect that (1) it is a pointer sub-region with size 8 bytes and (2) it should be rewired during copy in Phase 2.

At the end of Phase 1, the queue sub_regions contains all encountered and resolved at run time sub-regions in enclave memory. Listing 2 shows how these sub-regions are copied into untrusted shared memory

```
                                              Manifest file (in TOML syntax)

     // DRM_IOCTL_VERSION ioctl argument      sgx.ioctl_structs.drm_version = [
     struct drm_version {                       { size=3, unit=4, type="in" },
             int version_major;                 { size=1, unit=4, type="none" },  # padding
             int version_minor;                 { size=8, type="inout", name="name_len" },
             int version_patchlevel;            { ptr=[ {size="name_len", type="in",
             __kernel_size_t name_len;                   adjust=1} ] },           # NUL byte
             char __user *name;                 { size=8, type="inout", name="date_len" },
             __kernel_size_t date_len;          { ptr=[ {size="date_len", type="in",
             char __user *date;                          adjust=1} ] },           # NUL byte
             __kernel_size_t desc_len;          { size=8, type="inout", name="desc_len" },
             char __user *desc;                 { ptr=[ {size="desc_len", type="in",
     };                                                  adjust=1} ] }            # NUL byte
                                              ]
```

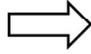

**Figure 6.** DRM_IOCTL_VERSION *ioctl* data structure as defined in the Linux DRM subsystem with the corresponding data structure in TOML syntax.

```
                                                 Manifest file (in TOML syntax)

     #define I915_CONTEXT_PARAM_BAN_PERIOD 0x1    sgx.ioctl_structs.drm_i915_gem_context_param = [
     #define ...                                    {size=4, type="out", name = "ctx_id"},
     #define I915_CONTEXT_PARAM_SSEU       0x7      {size=4, type="inout", name="size"},
     #define ...                                    {size=8, type="out", name="param"},
                                                    # int64 param I915_CONTEXT_PARAM_BAN_PERIOD
     // DRM_IOCTL_I915_GEM_CONTEXT_GETPARAM /       {onlyif="param == 0x1", size=8, type="inout"},
     // DRM_IOCTL_I915_GEM_CONTEXT_SETPARAM          ...
     // ioctl argument ("value" semantics           # ptr param I915_CONTEXT_PARAM_SSEU – pointer
     // depends on "param")                         # to nested drm_i915_gem_context_param_sseu
     struct drm_i915_gem_context_param {            {onlyif="param == 0x7", ptr=[{size=32,
             __u32 ctx_id;                                                        type="inout"}]},
             __u32 size;                            ...
             __u64 param;
             __u64 value;
     };
```

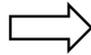

**Figure 7.** DRM_IOCTL_I915_GEM_CONTEXT_{GET/SET}PARAM *ioctl* data structure as defined in the Linux DRM subsystem with the corresponding data structure in TOML syntax.

in Phase 2. Note how the "`type`" parameter is used to copy sub-regions selectively; also note that sub-regions cannot be simply skipped but instead untrusted copies must be filled with zeros (to keep correct offsets). Finally, "`ptr`" sub-regions must be rewired to point to correct untrusted copies of pointed-to memory regions.

At the end of Phase 2, the *ioctl* data structure is fully copied to untrusted memory. At this point, the actual host-level *ioctl* system call is executed, with the `arg` argument pointing to the root of this copy. This system call may update some fields in this *ioctl* data structure copy. Listing 3 shows how the updated sub-regions are (selectively) copied back into enclave memory in Phase 3. Note that the sub-regions that must not be updated are "jumped over" in enclave memory (to keep correct offsets).

At the end of Phase 3, the complete *ioctl* emulation with deep copy is finished.

## 5. EVALUATION AND EXAMPLES

We initially added *mmap* and *ioctl* support in Gramine for GPU enablement. In particular, we targeted applications written against the Intel® Media SDK framework [12]. Media SDK uses the Direct Rendering Manager (DRM) standardized set of GPU-related *ioctl* requests [11]. Therefore, we enabled 32 DRM_IOCTL requests used by Intel® Media SDK. To access "device" pseudo-files in Linux, we added GPU-related "/dev" paths in the manifest file.

Figures 6 and 7 show two example DRM_IOCTL requests as we enabled them in the Gramine manifest (in TOML syntax). The first example is used by the application to retrieve information about the DRM version on the host. Thus, all fields are declared with type "`in`" (don't copy from enclave to untrusted memory, but copy back from untrusted to enclave memory). Some fields can be coalesced together in one sub-region – see how `version_major`, `version_minor` and `version_patchlevel` are defined as one sub-region of three 4-byte integers. Also note the dummy padding sub-region: the `drm_version` struct must be aligned to 8 bytes, thus there are 4 bytes after three 4-byte integers to pad this struct part to 16 bytes (C compilers add this padding transparently in resulting object code, but our TOML syntax must define these paddings explicitly). Finally note the use of the "`adjust`" keyword: each of the three strings is defined with length not size (i.e., `name_len`, `date_len`, `desc_len` do not count the NUL byte). Thus, our TOML syntax explicitly adds the adjustment to also copy this NUL byte.

The second example (in Figure 7) shows a more complicated *ioctl* data structure. This data structure has a common header consisting of `ctx_id`, `size` and `param` integers, but the rest layout is dependent on the value in `param`. For example, if `param` is equal to 1, then `value` is treated as an 8-byte integer that denotes the "ban period" for the GPU. As another example, if `param` is equal to 7, then `value` is treated as a pointer to a nested data structure containing Slice/Subslice/EU (SSEU) configuration (this nested data structure has a fixed size of 32 bytes). To choose at run-time among several `param`-conditioned layouts, we use the "`onlyif`" keyword.

After adding 32 *ioctl* requests with their corresponding data structures in the Gramine manifest file, we were able to execute unmodified Intel® Media SDK encode/decode/transcode workloads. Without going into evaluation details, we briefly note that we were able to achieve 95% throughput of the original, non-SGX transcode application. This good performance is partly due to the fact that *ioctl* requests do not dominate the transcode runtime (it is dominated by the GPU execution itself), thus our deep copy doesn't become a bottleneck. And partly due to several caching optimizations that we added to the basic implementation in Listing 1.

## 6. CAVEATS AND LIMITATIONS

While enabling real-world Intel® Media SDK workloads, we encountered several problematic cases that we had to work around in Gramine. These problematic cases highlight the limitations and caveats of transparent device-specific *mmap* and *ioctl* support in completely unmodified applications.

**Caveat 1.** All memory regions that will be shared with the host device must be allocated via device-backed mmap.

At *mmap* invocation time, Gramine must choose whether the memory region will be allocated in enclave memory (anonymous and file-backed mappings) or in untrusted memory (device-backed mappings). Once the region is allocated in enclave memory, this region cannot be shared with the device. This is the consequence of the two-world (trusted enclave and untrusted rest) nature of the Intel® SGX technology.

However, this separation does not exist in normal, non-SGX environments: any previously allocated memory region can be shared with the device. For example, a data buffer can be allocated via anonymous mmap and only later marked as shared with the GPU; we have seen this pattern in AI/ML frameworks that choose whether to execute computations on the CPU or to offload them on the GPU at runtime, after all data is allocated. This pattern would **not** work in Gramine with SGX. Frameworks and applications that use this pattern would need to be rewritten. Fortunately, Intel® Media SDK does not exhibit this pattern and thus does not display this caveat.

**Caveat 2.** Polling on the *ioctl* data structure allocated inside the enclave and intended for host-device updates hangs the application.

Figure 8 illustrates this caveat. The application allocates an *ioctl* data structure `buf_for_gpu` and executes the *ioctl* request. But this *ioctl* request has peculiar semantics: it returns immediately, and the GPU notifies on its completion by updating a field of the data structure. The application waits for this field to be updated by periodically reading its value (this is called *polling*). However, the `completion_byte` pointer points to the original data structure in enclave memory, but the GPU updates the field in the copied data structure in untrusted memory. The SGX

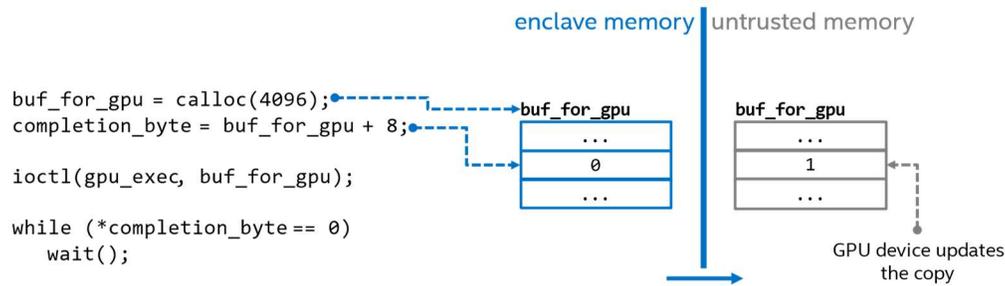

**Figure 8.** Problem of polling on the *ioctl* data structure: GPU updates the untrusted-memory copy but the SGX application waits for update in enclave-memory original data structure.

application never learns about the update because it accesses the wrong memory cell, and it hangs.

Unfortunately, there is no simple fix for this polling issue. First, the GPU device cannot be asked to update the in-enclave field because Intel® SGX technology prohibits this. Second, Gramine cannot allocate `buf_for_gpu` immediately in untrusted memory because `calloc()` memory-allocation routine provides no context/hint where the buffer should be allocated (so Gramine conservatively allocates it in enclave memory). Third, the application's `completion_byte` pointer cannot be transparently rewired by Gramine to point to the untrusted-memory copy because in a compiled application binary, reading through the `completion_byte` pointer becomes indistinguishable from normal memory accesses. In other words, Gramine has no hint from the application to "hook onto".

The only way to circumvent this caveat is to change the application code: either not use the code with this polling pattern at all, or rewrite the code such that it allocates the buffer in untrusted memory. Fortunately, we ran into only one such place in Intel® Media SDK. We worked around this by replacing the offending *ioctl* request with another, similar *ioctl* request that doesn't have this polling pattern.

**Caveat 3.** Our deep copy implementation prohibits allocating new objects in enclave memory during *ioctl* request emulation.

It is currently impossible to force Gramine to allocate objects inside the enclave memory during ioctl request emulation. For example, it would be impossible for the untrusted runtime or the device to add new items to the `ops` linked list in Figure 1. This is not a limitation of our deep copy approach but rather a conscious design decision: allowing new objects coming from untrusted environment decreases security. Thus, to close this attack vector, we chose to prohibit copying new objects from untrusted memory into enclave memory. Fortunately, we encountered no *ioctl* requests in Intel® Media SDK that would require such functionality. All the encountered *ioctl* requests allocating objects/buffers in untrusted memory do not need to copy these objects inside the enclave, instead only the pointers to these objects are copied inside the enclave.

**Caveat 4.** Some *ioctl* requests require special handling in Gramine: creating new file descriptors (FDs), sanitizing returned *ioctl* data structures, etc.

One troublesome feature of *ioctl* requests is that they may have side effects, affecting the internal Gramine state. For example, the DRM_IOCTL_I915_GEM_EXECBUFFER2_WR *ioctl* request creates a "fence" file descriptor to poll on inside the enclave, waiting for the GPU to finish. (Note that this polling is similar to Caveat 2, but much better for Gramine because there is an actual "fence" object to intercept and track rather than some opaque pointer.) Upon this ioctl execution, Gramine needs to re-create this "fence" file descriptor in its internal state and pass it back to the application, so that the application may poll on this file descriptor, and Gramine can track these poll attempts. Since this logic has nothing to do with deep copying and requires very specific logic in Gramine, we implement it directly in Gramine source code.

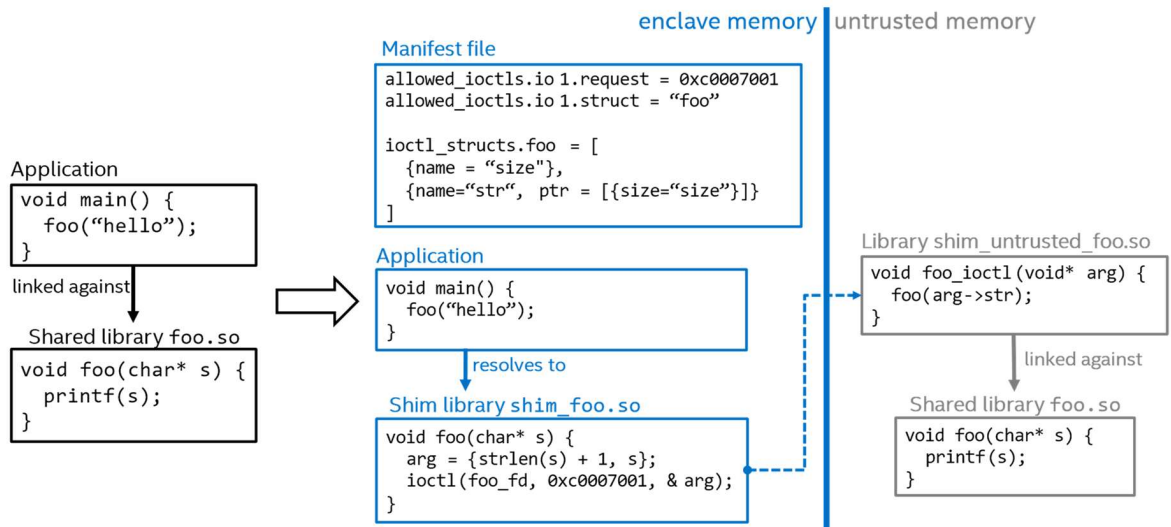

**Figure 9.** Re-purposing *ioctl* requests for slicing the SGX application. Here, the shared library `foo.so` is kept outside of the enclave, and two helper shim libraries `shim_foo.so` and `shim_untrusted_foo.so` marshal shared-library function calls by replacing them with *ioctl* requests.

We also encountered couple *ioctl* requests that required sophisticated sanitization of the returned values in *ioctl* data structures. In particular, DRM_IOCTL_I915_GETPARAM with the I915_PARAM_HAS_BSD2 parameter must return *false* in Gramine because Gramine doesn't implement Sys-V IPC primitives (implied when this parameter returns *true*). We pondered on whether to add a new TOML syntax keyword like **"force_to_value"** but opted against this: this sanitization is too rare and too specific to merit its own keyword. To date, we encountered only two *ioctl* requests requiring some form of sanitization.

Other than the above four caveats, we saw no other problems with enabling Intel® Media SDK workloads in Gramine and with our *mmap/ioctl* support.

## 7. USAGES BEYOND DEVICE COMMUNICATION

The work on *mmap/ioctl* support in Gramine was started with a specific use case in mind: enable GPU offload for Intel® Media SDK workloads. This support was designed with generality in mind – we plan to support other accelerators (TPUs, FPGAs, network cards, etc.) and other frameworks (OpenCL, CUDA, etc.). In theory, adding support for new accelerators and frameworks is only a matter of defining their corresponding *ioctl* requests in our proposed TOML syntax, and possibly working around the caveats mentioned in the previous section.

With time, we came to realize that our generic *ioctl* support and the deep copy implementation can be used beyond initially planned communication with host devices. At a high level, our work is akin to the Enclave Definition Language (EDL) in Intel® SGX SDK and to the marshalling libraries like Google Protocol Buffers [15, 21].

Therefore, we propose to repurpose the *ioctl* support in this work for generic OCALL implementations. In particular, we propose to slice the original application into two parts: the one that runs inside the enclave and the one that runs outside of it (in untrusted memory). The slicing happens at the level of shared libraries: this is a natural boundary between two components [23, 24]. To this end, we propose to transparently replace shared-library calls with *ioctl* requests: the in-enclave application packs library call's arguments in an *ioctl* data structure and executes the *ioctl* request, the untrusted runtime intercepts this ioctl request and forwards it to the (untrusted) shared library, the shared library finally updates the *ioctl* data structure, and control returns back to the application.

The original application doesn't need to be modified: it is enough to create a small shim library that will intercept shared-library calls and transform them into *ioctl* requests. The application is then re-linked against this shim library. Of course, the application must be also accompanied by the manifest file that defines the *ioctl* requests.

Figure 9 shows this concept on a simple example. The original application links against the library `foo.so` and calls its `foo()` function. We want to partition (slice) this original application into two components: the application itself runs inside the SGX enclave whereas the shared library runs outside of the SGX enclave. To achieve this, we create two helper shim libraries: `shim_foo.so` (runs inside the enclave) and `shim_untrusted_foo.so` (runs outside of the enclave). We also define the `foo` *ioctl* data structure to copy outside of the enclave.

At run-time, instead of loading `foo.so`, we instruct Gramine to load `shim_foo.so` which will intercept calls to `foo()`. The interception function wraps the single argument `s` into an *ioctl* data structure and invokes *ioctl* emulation. This emulation exits the enclave and lands in `foo_ioctl()` function of another shim library `shim_untrusted_foo.so`. At this point, the argument `s` is de-wrapped from the copied *ioctl* data structure and forwarded to the actual `foo()` implementation.

Notice how the original application and the original shared library are not modified at all. This partitioning implementation is similar in spirit to Remote Procedure Calls (RPCs) used for example in Java RMI [25] and in Google's gRPC [26], but doesn't force refactoring/rebuilding of the original application.

We would like to stress that this splitting does not provide any security guarantees on the untrusted shared memory and the *ioctl* communication channel. As with device support, we consider this a non-goal of this work: the application itself must be designed in such a way as to better protect from attacks on the untrusted memory and on the *ioctl* data structures.

## 8. CONCLUSION

In this paper, we demonstrated our device-backed *mmap* support and generic *ioctl* support in Gramine. The *mmap* support allows to create memory regions in untrusted memory shared between the SGX enclave and the host device. The *ioctl* support allows to create a communication channel between the SGX enclave and the host device, and is based on two key ingredients: a new TOML syntax to describe *ioctl* requests and their data structures, and a deep copy algorithm to selectively copy *ioctl* data structures in and out of the enclave. We demonstrated its applicability for GPU offload on Intel® Media SDK workloads. The proposed *mmap/ioctl* support is generic enough to be applicable to our offload devices like TPUs, FPGAs, etc. It also paves a way for application slicing into trusted and untrusted parts.

This work (at least partially) is planned to be upstreamed.

## 9. ACKNOWLEDGEMENTS

We would like to acknowledge Michał Kowalczyk and Anjo Lucas Vahldiek-oberwagner for their valuable feedback and suggestions.

## 10. NOTICES & DISCLAIMERS